\documentclass[conference]{IEEEtran}
\IEEEoverridecommandlockouts
\usepackage{cite}
\usepackage{amsmath,amssymb,amsfonts}
\pdfoutput=1
\usepackage{graphicx}
\usepackage{textcomp}
\usepackage{xcolor}
\usepackage{caption}
\usepackage{subcaption}
\usepackage{url}

\usepackage{algorithmic}
\usepackage[linesnumbered,ruled,vlined]{algorithm2e}
\usepackage{color, soul, colortbl}
\usepackage{xcolor}

\SetKwFor{ParFor}{for}{do in parallel}{}

\usepackage{tikz}
\newcommand*\circled[1]{\tikz[baseline=(char.base)]{
    \node[shape=circle,draw,inner sep=0.5pt] (char) {\small#1};}}

\PassOptionsToPackage{hyphens}{url}\usepackage{hyperref}

\def\BibTeX{{\rm B\kern-.05em{\sc i\kern-.025em b}\kern-.08em
    T\kern-.1667em\lower.7ex\hbox{E}\kern-.125emX}}

\begin{document}

\title{FSHMEM: Supporting Partitioned Global Address Space on FPGAs for Large-Scale Hardware Acceleration Infrastructure
\thanks{
This research was supported by the MSIT (Ministry of Science and ICT), Korea, under the ITRC (Information Technology Research Center) support program (IITP-2020-0-01847) supervised by the IITP (Institute of Information \& Communications Technology Planning \& Evaluation).
}
}


\author{
    \IEEEauthorblockN{
        Yashael Faith Arthanto
    }
    \IEEEauthorblockA{
        \textit{School of Electrical Engineering} \\
        \textit{KAIST}\\
        Daejeon, Republic of Korea \\
        yashael.faith@kaist.ac.kr
        \vspace{-0.5in}
    }
    \and
    \IEEEauthorblockN{
        David Ojika
    }
    \IEEEauthorblockA{
        \textit{Flapmax} \\
        Austin, USA \\
        dave@flapmax.com
        \vspace{-0.5in}
    }
    \and
    \IEEEauthorblockN{
        Joo-Young Kim
    }
    \IEEEauthorblockA{
        \textit{School of Electrical Engineering} \\
        \textit{KAIST}\\
        Daejeon, Republic of Korea \\
        jooyoung1203@kaist.ac.kr
        \vspace{-0.5in}
    }
}

\maketitle

\begin{abstract}
By providing highly efficient one-sided communication with globally shared memory space, Partitioned Global Address Space (PGAS) has become one of the most promising parallel computing models in high-performance computing (HPC).
Meanwhile, FPGA is getting attention as an alternative compute platform for HPC systems with the benefit of custom computing and design flexibility. 
However, the exploration of PGAS has not been conducted on FPGAs, unlike the traditional message passing interface.
This paper proposes FSHMEM, a software/hardware framework that enables the PGAS programming model on FPGAs.
We implement the core functions of GASNet specification on FPGA for native PGAS integration in hardware, while its programming interface is designed to be highly compatible with legacy software. 
Our experiments show that FSHMEM achieves the peak bandwidth of 3813 MB/s, which is more than 95\% of the theoretical maximum, outperforming the prior works by 9.5$\times$. It records 0.35$us$ and 0.59$us$ latency for remote write and read operations, respectively.
Finally, we conduct a case study on the two Intel D5005 FPGA nodes integrating Intel's deep learning accelerator.
The two-node system programmed by FSHMEM achieves 1.94$\times$ and 1.98$\times$ speedup for matrix multiplication and convolution operation, respectively, showing its scalability potential for HPC infrastructure.
\end{abstract}


\vspace{-0.35in}
\textbf{}\section{Introduction}
High Performance Computing (HPC) employs computer clusters to solve advanced computational problems, primarily centered around scientific applications including molecular modeling~\cite{shaw-molecular-hpc}, weather modeling~\cite{jordan-weather-hpc}, and genomics~\cite{simpson-genomics-hpc}.
Recently, as artificial intelligence (AI) and machine learning (ML) technology are transforming major industries with highly beneficial applications such as image captioning \cite{vinyals2016show, cornia2020meshed}, virtual assistant \cite{Capes2017}, stock market prediction \cite{Mehtab2021stock}, and self-driving cars \cite{tesla2020micro}, HPC systems have evolved to accommodate AI workloads\cite{coates2013deep, kalamkar2020optimizing}. 
Consisting of hundred-thousands of CPU-GPU  nodes, a state-of-the-art HPC infrastructure performs several hundreds of peta floating-point operations per second (PFLOPS) \cite{monroe2020fugaku, hines2018stepping}.
\textit{Parallel programming model} is the key to run large-scale AI services on distributed computing nodes.

\begin{figure}[b]
  \vspace{-\baselineskip}
  \vspace{-0.05in}
  \centering
  \includegraphics[width=0.9\linewidth]{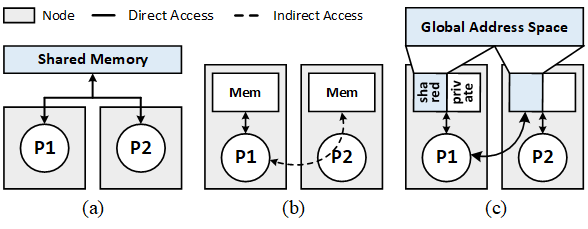}
  \vspace{-0.07in}
  \caption{Three parallel programming models: (a) Shared Memory, (b) Message Passing, and (c) PGAS}
  \label{prog_model}
  \vspace{-0.05in}
\end{figure}

Figure~\ref{prog_model} shows three main parallel programming models: Shared Memory, Message Passing, and Partitioned
Global Address Space (PGAS). 
In the shared memory model, multiple processes can directly access the shared memory space. On the other hand, the message passing model only allows the processes to exchange messages based on the Message Passing Interface (MPI)~\cite{mpi40}.
MPI has become popular in HPC as it enables data communication between remote nodes with the same inter-process mechanism. 
Since version 2, it also supports one-sided data communication operations.
PGAS uses one-sided data communication with the concept of partitioned but globally shared memory.
Leveraging the benefit of the shared memory model, each node has direct access to another node's memory space without interfering with the corresponding process. 
This is possible because all the nodes in the network form a single global address space, in which each node has its own part. In addition, each node has a private memory for its local processing. 
With this concept, PGAS brings a few advantages to parallel programming.
First, its shared view of memory simplifies parallel programming, inspired by the shared memory paradigm. 
Second, PGAS's one-sided communication \cite{yelick2007prod} that does not interfere with the remote process reduces the communication overhead.
Third, the clear decoupling of private and shared memory gives programmers a better perspective on utilizing locality.

Meanwhile, the HPC community has a surge of interest in adopting an alternative acceleration platform beyond GPU. FPGA is one strong candidate due to the flexibility in design, proving its feasibility in production environments~\cite{brainwave2018, mango, cygnus}.
For the use of FPGA in HPC, several works have implemented MPI on FPGAs~\cite{xiong20-collectives, tmd-mpi, patel06-scalable, ly09-challenge, axel}. However, research on PGAS support for FPGA is still preliminary~\cite{willenberg14-gasnet}, as the protocol is relatively new. 
In this paper, we present FSHMEM, a software/hardware framework to support PGAS on FPGAs, named after the shared memory library (SHMEM). Along with the trend of FPGA adoption in HPC, we believe that FSHMEM will play a fundamental role in building FPGA-based hardware infrastructure at scale. Our main contributions are as follows:
\begin{itemize}
\item FSHMEM implements core functions of GASNet protocol\cite{gasnet_1.8.1} on FPGA to enable native PGAS integration in hardware. It also provides an easy-to-use software interface for high adaptability.
\item We benchmark the FSHMEM's bandwidth and latency performance for Remote Direct Memory Access (RDMA) and compare it against the previous results. Our maximum bandwidth of 3813 MB/s outperforms prior works by 9.5$\times$ with an average latency of 0.47$\mu s$.
\item We build a practical multi-FPGA acceleration system using FSHMEM and Intel's Deep Learning Accelerator (DLA) as computing core. We parallel-program the AI computations and evaluate the performance to show the framework's potential for performance scaling.
\end{itemize}
\vspace{-0.1in}
\section{Background}
\vspace{-0.05in}
\subsection{GASNet}
\vspace{-0.02in}
Global Address Space Networking (GASNet) is a language-independent networking middleware that describes PGAS's one-sided communication interface.
Its interface is built around Active Message (AM) protocol. In AM, each message head specifies the address of a handler function that will be called upon the message's arrival, and each message body provides the arguments for the function along with the data to be transferred \cite{am}.
GASNet does not provide function implementations because it may vary among network interfaces or devices, but it must support one-way data communication. 
Recently, GASNet-EX, an upgraded version of GASNet, was also introduced\cite{bonachea2018gasnet}. It achieves an average of 1.77$\mu s$ latency and saturates to maximum bandwidth at 4-8KB, which are 1.05-5$\times$ faster in latency and 4$\times$ faster to saturation than MPI.

\vspace{-0.05in}
\subsection{MPI on FPGA}
\vspace{-0.02in}
TMD-MPI \cite{tmd-mpi2} implemented MPI on FPGA by creating a software library for embedded processors and TMD-Message Passing Engine (TMD-MPE) for hardware kernels.
This engine brings MPI functionality to a hardware kernel by handling MPI's protocol and packet generation. They perform bandwidth and latency benchmarks under a 2-rank system and discover a maximum bandwidth at 400 MB/s, achieving 75\% of its peak bandwidth.
Ziavras \textit{et al.} implemented one-sided MPI primitives on embedded FPGA~\cite{ziavras-mpi}. Tested with two FPGAs on a single board, their implementation reaches 141 MB/s, or 70.6\% of the peak bandwidth.
Other works~\cite{xiong20-collectives, haghi20fpga} try to improve the system bottleneck caused by extensive use of collective communications by offloading the data processing algorithms such as \emph{reduce, allReduce, bcast} to the FPGAs within network switches. They attain the latency speedup of 3.9$\times$ on average.

\vspace{-0.05in}
\subsection{PGAS on FPGA}
\vspace{-0.02in}
THe GASNet (Toronto Heterogeneous GASNet)\cite{willenberg14-gasnet} is the latest work that implemented GASNet on FPGA by introducing Global Address Space Core (GASCore), an RDMA core, and Programmable Active Message Sequencer (PAMS).
PAMS is responsible for interfacing the hardware kernel to the GASCore by handling messages and synchronization. 
The GASCore can be programmed directly through software when used by embedded processors.
Meanwhile, hardware kernels program the GASCore through PAMS, containing a code memory for GASNet operations.
In addition, they suggest a software stack to transform an application's compute portion to hardware kernels and load the GASNet portion into PAMS.

\vspace{-0.05in}
\subsection{Multi-FPGA Infrastructure for HPC}
\vspace{-0.02in}
Axel\cite{axel} showcased an early adoption of FPGAs in HPC infrastructure, in which each node includes CPU, GPU, and FPGA through a PCIe switch. 
It uses a Map-Reduce framework to accelerate N-Body Simulation using heterogeneous resources achieving an impressive 22.7$\times$ performance improvement. 
However, Axel does not scale out well, getting only 4.4$\times$ improvement from 1 to 16 nodes because of the all-to-all functions that do not scale linearly with the number of nodes.
On the other hand, Microsoft's BrainWave platform\cite{brainwave2018} deployed FPGAs in the data center at scale to provide real-time AI services. In this architecture, FPGA accelerators are integrated between the CPUs and the top-of-rack switch, allowing them to communicate with many servers in the data center. Each FPGA contains a specialized AI accelerator for low latency inference services.

Cygnus~\cite{cygnus} is a heterogeneous supercomputer utilizing FPGAs in 40\% of their compute nodes. These FPGAs are additionally interconnected in a 2D-torus Infiniband network. Cygnus achieved 2.4 PFLOPS for double-precision, while the FPGAs contributed only 0.6 PFLOPS for single-precision.
\begin{figure}
  \centering
  \includegraphics[width=0.92\linewidth]{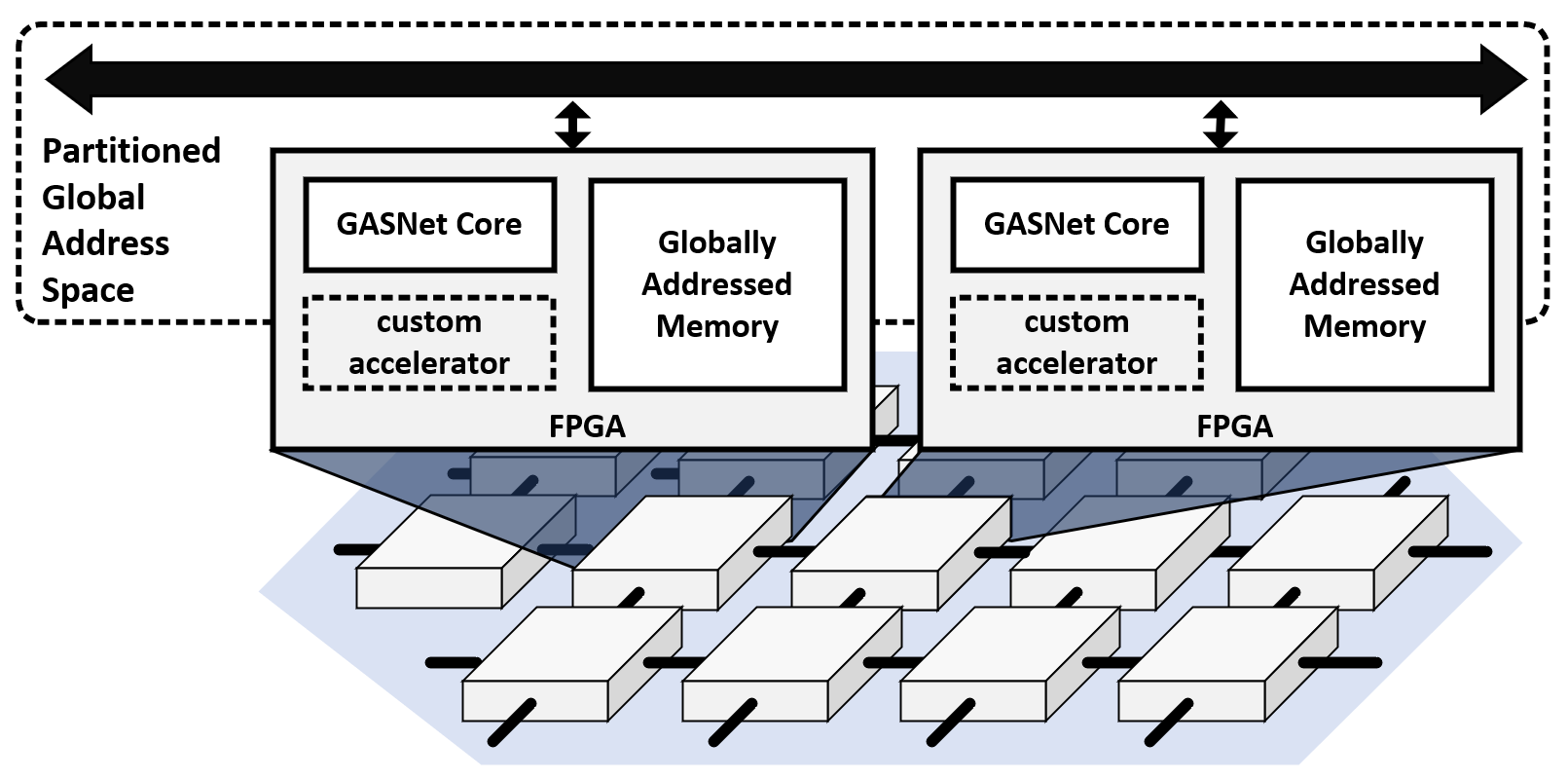}
  \vspace{-0.02in}
  \caption{FSHMEM-based hardware acceleration infrastructure}
  \label{high-level}
  \vspace{-\baselineskip}
  \vspace{-0.05in}
\end{figure}

\begin{figure*}
  \vspace{-0.1in}
  \centering
  \includegraphics[width=\linewidth]{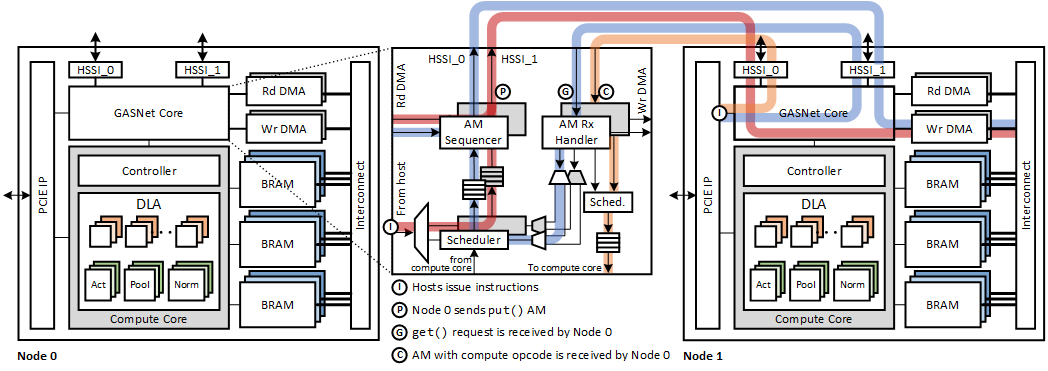}
  \caption{FSHMEM node architecture with dataflow for \emph{gasnet\_put} (red), \emph{gasnet\_get} (blue) and \emph{gasnet\_AMRequest*} (orange)}
  \label{arch}
  \vspace{-0.1in}
\end{figure*}

\section{FSHMEM: Inter-FPGA Infrastructure for PGAS Programming Model}
This work proposes FSHMEM, an infrastructural framework that enables PGAS on FPGAs by implementing the GASNet interface in hardware and supporting the corresponding API in software. 
Figure~\ref{high-level} shows the conceptual diagram of an FSHMEM-based hardware acceleration infrastructure made of fabrics of FPGAs. 
FSHMEM enables the PGAS parallel computing model on a pool of FPGAs by interconnecting them through the GASNet interface. 
Each FPGA node in this infrastructure includes the GASNet core, a globally addressed memory space, and a custom accelerator. Note that the hardware infrastructure can adopt any network topology, while the diagram shows an example mesh topology.

The FSHMEM based infrastructure facilitates mainly three benefits. 
First, FSHMEM provides low latency, direct FPGA-to-FPGA communication by implementing GASNet's lightweight interface on high-speed transceiver links. 
Second, FSHMEM is highly flexible with globally shared memory space and local memories. It allows users to deliberately manage memory transfer and internal data flow using GASNet's memory-to-memory functions and custom hardware function handlers.
Third, FSHMEM's software API is highly compatible with the existing PGAS frameworks so that many legacy HPC applications can be easily adapted to use FSHMEM.
In the following subsections, we will describe the GASNet core, an essential hardware module to implement the GASNet protocol on FPGA, the custom accelerator's interface to the GASNet core, and the software interface.

\vspace{-0.05in}
\subsection{GASNet Core}

GASNet core implements the fundamental communication protocol of the GASNet specification, which is the AM interface.
There are two key features to consider when implementing AM on hardware:

\begin{itemize}
    \item AM invokes a handler function on the destination node that may compute the data or reply with the requested data. This is done by passing a handler function pointer in a software implementation. Instead, the GASNet core directly passes the function opcode.
    \item AM carries arguments for the function to be invoked. Arguments can be integers or data payloads to be stored in the destination address. Therefore, to provide such functionality, the GASNet core must contain an RDMA.
\end{itemize}

\begin{table}[t]
    \centering
    \caption{Implemented Functions on GASNet Core}
    \label{tab:fn}
    \def\arraystretch{1.4}%
    \begin{tabular}{ p{0.3\linewidth} p{0.6\linewidth} }
        \hline
        Function & Description \\
        \hline
        \emph{gasnet\_AMRequest*()} & Send messages to destination node \\
        \emph{(Short/Medium/Long)} & {} \\
        \hline
        \emph{gasnet\_AMReply*()} & Reply messages to requesting node \\ 
        \emph{(Short/Medium/Long)} & {} \\
        \hline
        \emph{gasnet\_put()} & Store data in the target node \\
        \hline
        \emph{gasnet\_get()} & Request data from the desired node \\
        \hline
    \end{tabular}
    \vspace{-0.2in}
\end{table}

Table \ref{tab:fn} shows the list of GASNet functions that GASNet core currently supports on the FPGA: \emph{gasnet\_AMRequest}, \emph{gasnet\_AMReply}, \emph{gasnet\_put}, and \emph{gasnet\_get}. 
Other functions from the specifications such as job controls, job environments, and barrier functions are implemented on the software side. Meanwhile, atomicity control of the handler function is natively supported by hardware.

As part of the specification, AM request functions have three variants based on the length of arguments: short, medium, and long.  
The short message does not send a payload to the destination, making it generally suitable for configuration update functions.
Both medium and long messages include payload, but medium messages go to the local memory address while long messages go to the globally shared address space.
GASNet's AM reply functions are essentially the same as the request functions, except they can only reply to the requesting node.
\emph{gasnet\_put} and \emph{gasnet\_get} function, which comes from the GASNet extended API, can be implemented using the request and reply functions.
For example, we use the AM request function that invokes the PUT and GET handler for \emph{gasnet\_put} and \emph{gasnet\_get} function, respectively.
Note that the GET handler will invoke an AM reply function.
Another necessary handler is for the compute core.

Figure~\ref{arch} illustrates the FSHMEM's node architecture, mainly consisting of the host interface (PCIe), GASNet core, and its underlying memory and accelerator subsystems.
The GASNet core is composed of two sets of AM sequencer, AM receiver handler, and schedulers with FIFOs. Each set corresponds to the High-Speed Serial Interface (HSSI) port for inter-FPGA communication. 
In this module, the AM sequencer forms the active message by generating the header and reading the message body from the memories via DMA. 
Since the requests can come from multiple sources, e.g., host, compute core, or a remote node, the scheduler is necessary. 
The AM receiver handler writes the incoming data to the memories.

The figure also shows the fundamental data flows between two FSHMEM nodes: \emph{gasnet\_put} in red, \emph{gasnet\_get} in blue, and \emph{gasnet\_AMRequest} in orange. For the \emph{gasnet\_put} operation, which is a remote write from Node 0 to Node 1, the host of Node 0 initially issues the \emph{gasnet\_put} command (\circled{I}). Going through the scheduler and FIFO, it arrives at the AM sequencer. 
The sequencer then fetches the necessary data using the read DMA and sends the formed message to Node 1 (\circled{P}).
In Node 1, the AM receive handler checks the opcode of the received message, which should be a PUT opcode, and uses write DMA to store the payload to the destination address.

Let us assume that Node 1 performs \emph{gasnet\_get} operation this time, which is a remote read from Node 0. Similar to the \emph{gasnet\_put} case, the host of Node 1 issues the \emph{gasnet\_get} command, yet without carrying any payload.
Upon the arrival of the GET request in Node 0 (\circled{G}), the AM receiver handler immediately issues a PUT reply command and forwards it to the scheduler. After that, the execution is exactly the same as the PUT operation described above. As a result, the requested data are read and packed by the AM sequencer and are sent to Node 1, which accomplishes the GET command.
The orange color in the figure highlights the dataflow of a \emph{gasnet\_AMRequest*} function, especially with the case that carries a compute opcode without data payloads. 
Upon receiving this type of message (\circled{C}), the AM receiver handler sends the function's arguments to the compute command scheduler to get it queued for the compute core's execution. 
If the received message carries the data payload, the AM receiver handler writes it into the memory before instructing the compute core to execute.


Due to the simple and optimized design, the GASNet core's logic usage is extremely low, around 0.2\% logic for two HSSI ports on Intel Stratix 10. Its logic size will increase with the number of available HSSI ports in the FPGA. This low design overhead allows the compute core to utilize most of the device's resources for high application performance. 
However, as the GASNet core is not designed for any specific network topology, it may need a router for an extensive network setting.

\subsection{Deep Learning Accelerator}
For the primary compute core of this work, we customize the Intel DLA~\cite{intel-dla} for major AI computations such as convolution and matrix multiplication. 
It uses a 1-dimensional systolic array architecture to accelerate these computations with high flexibility. 
Although optimized for convolution, the DLA can also perform matrix-vector and matrix-matrix multiplication by properly mapping the data.
It is highly flexible as the computation types and tensor sizes are exposed as arguments. The GASNet's active message can easily instruct the DLA via its handler interface by passing a few arguments.

A typical interaction between the host and DLA for parallel execution is a repetition of compute command, acknowledgment, and PUT command. With a powerful compute core like the DLA, this workflow requires an extra host intervention and a considerable communication bandwidth if the PUT command is performed upon the final result data.
Therefore, we devise the Automatic Result Transfer (ART) mechanism to let the DLA initiate the transfer.
Instead of sending a large-sized message at the end of the computation, ART splits it into several smaller-sized messages in the middle of the computation. ART enables this by issuing a PUT command for every $N$ valid result, in which $N$ is configurable. Thus, ART hides the communication latency with the computation execution time while removing extra host interventions.

\vspace{-0.05in}
\subsection{Software Interface}

\begin{figure}
  \centering
  \includegraphics[width=0.9\linewidth]{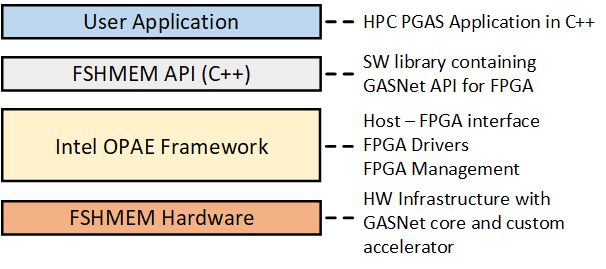}
  \vspace{-0.1in}
  \caption{FSHMEM software stack}
  \label{sw-stack}
  \vspace{-\baselineskip}
  \vspace{-0.1in}
\end{figure}

FSHMEM supports GASNet API for compatibility with legacy HPC applications. Figure~\ref{sw-stack} shows FSHMEM's entire software stack that starts from the user's PGAS-based HPC application. The application is programmed using the GASNet-compatible FSHMEM API in C++. The stack utilizes Intel's Open Programmable Acceleration Engine (OPAE) library~\cite{intel-opae} for host communication, device driver, and FPGA management. Although FSHMEM is currently based on Intel's FPGA platform, we can extend it to Xilinx's FPGA platform by integrating Xilinx Runtime (XRT) library~\cite{xilinx-xrt}. On the bottom of the stack, FSHMEM provides a hardware layer that implements remote data communications across devices. Each FSHMEM device instantiates the GASNet core module that works alongside the software interface.
Compared to THe GASNet, our FSHMEM API directly commands the GASNet core to initiate data transfer without needing to compile or translate code into dedicated instructions. In this way, FSHMEM promotes a 'plug-and-play' notion of using FPGA with GASNet API.

\section{Communication Performance Results}

\subsection{Methodology}
We use Intel Acceleration Stack 2.0.1 framework for logic synthesis and layout and implement the architecture on Intel's D5005 Programmable Acceleration Card (PAC). Intel D5005 PAC is a high-performance PCIe-attached FPGA acceleration card that includes Stratix 10 SX FPGA (part number: 1SX280HN2F43E2VG) with 32GB DDR memory and 2 QSFP+ network interfaces.
For experiments, we build a prototype machine that harnesses an Intel CPU as the host and two PACs as the FSHMEM devices. The host CPU drives the testing/application program using FSHMEM API, while both PACs, interconnected via QSFP+ cables in a ring fashion, are responsible for actual execution.
We profile FSHMEM's PUT and GET active messages via \emph{gasnet\_put} and \emph{gasnet\_get} function by measuring the read/write bandwidth and latency between the two FPGAs.
For accurate measurement, we add a hardware performance counter to measure the time taken from when a command is given until the corresponding message is returned. 

\vspace{-0.05in}
\subsection{FPGA Resource Utilization}
\begin{table}
    \centering
    \caption{FPGA Resource Utilization}
    \label{tab:util}
    \def\arraystretch{1.4}%
    \begin{tabular}{ p{0.2\linewidth} p{0.25\linewidth} p{0.15\linewidth} p{0.2\linewidth} }
        \hline
        Module & LUT + Register & BRAM & DSP \\
        \hline
        GASNet core & 1995.3 (0.21\%) & 17 (0.15\%) & 0 (0\%) \\
        DLA & 102276 (10.96\%) & 8 (0.07\%) & 1409 (24.46\%) \\
        \hline
    \end{tabular}
    \vspace{-0.2in}
\end{table}

Table \ref{tab:util} shows the resource utilization result of the GASNet core and DLA implemented on Intel D5005 PAC at 250 MHz operating frequency.
The GASNet core takes minimal logic resources to implement, making it suitable for use with deep learning accelerators or other compute-demanding accelerators.
The DLA used for our experiments contains 16$\times$8 PEs, utilizing a quarter of the available DSPs.

\vspace{-0.05in}
\subsection{Communication Bandwidth}

\begin{figure}[t]
  \vspace{-0.2in}
  \centering
  \includegraphics[width=0.9\linewidth]{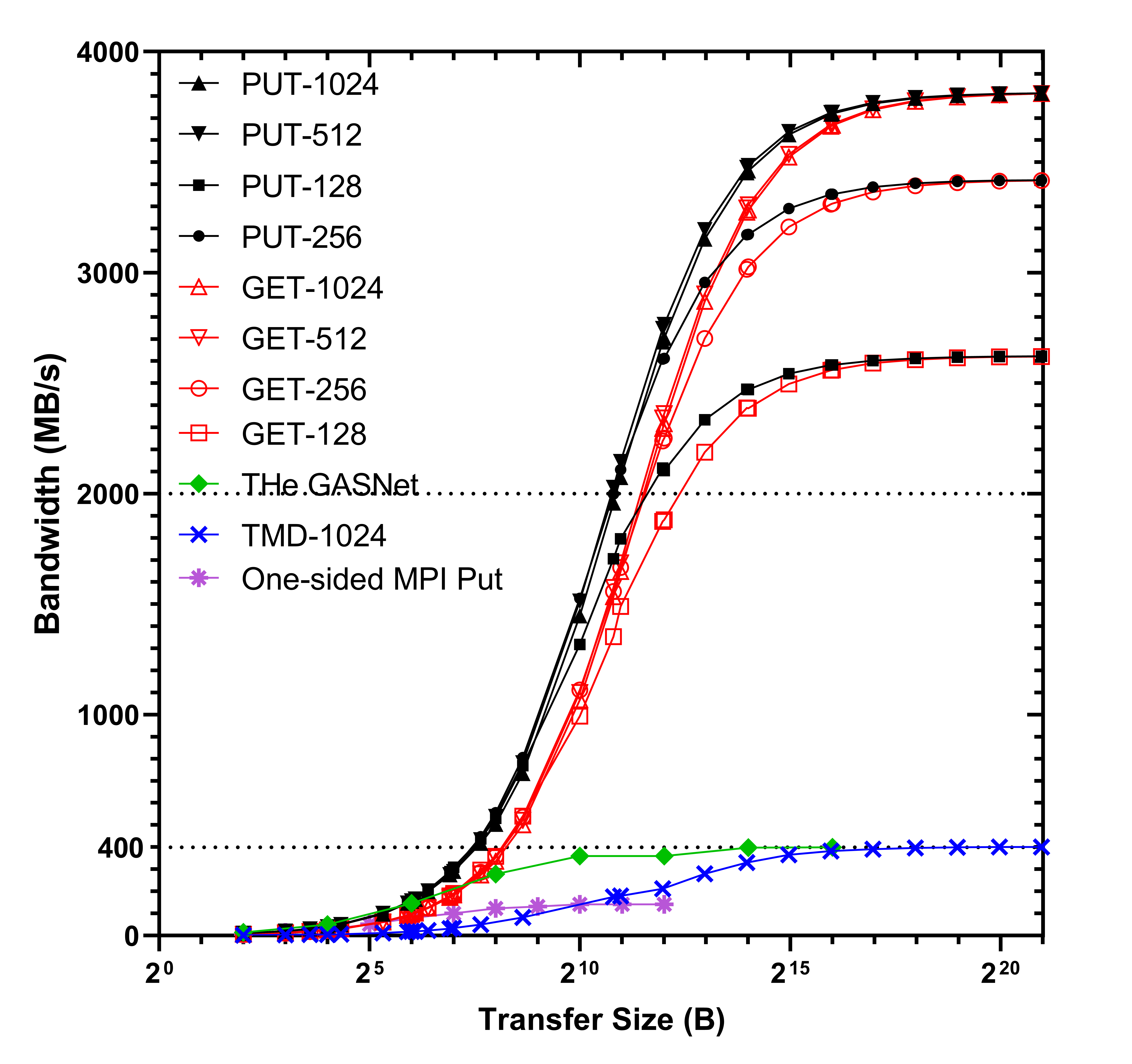}
  \vspace{-0.1in}
  \caption{Communication bandwidth measurement results}
  \label{fig:bw-results}
  \vspace{-0.1in}
\end{figure}

Figure~\ref{fig:bw-results} shows the communication bandwidth measurement results of FSHMEM when the packet size varies among 128, 256, 512, and 1024 bytes, with increasing transfer size from 4 bytes to 2 MB. We measure both PUT and GET bandwidth 
and achieve a peak bandwidth of 3813 MB/s for the packet size of 512 and 1024 bytes, which is more than 95\% of the theoretical maximum bandwidth. 
The 128 and 256-byte packet size achieves a 2621 and 3419 MB/s peak bandwidth, which is 65\% and 85\% of the theoretical maximum, respectively.
Smaller packet sizes show lower peak bandwidths because they need to generate more packets for the same transfer size, causing network overhead and throughput degradation.

We have two observations in the bandwidth curves. First, FSHMEM's communication reaches the half-maximum at around 2KB. It saturates around the transfer size of 32KB, reaching 95\% of the peak bandwidth for all cases, telling us that we must transfer at least 32KB to fully utilize the FSHMEM's available bandwidth.
Second, we observe that GET functions (i.e., read operation) have lower bandwidth than PUT functions (i.e., write operation). Although the GET bandwidth is almost similar to PUT when the transfer size is large enough ($>$ 32KB), their gap tends to be large for small-to-medium-sized transfers. For example, the GET bandwidth is 20\% and 8\% lower than the PUT bandwidth at the transfer size of 2KB and 8KB, respectively. 
This is because the GET function consists of a short request message and a long reply message with data, while the PUT function only consists of a long message containing the data. 
Therefore, the GET function includes an additional short request message that does not contain any payload, causing a constant overhead regardless of the transfer size.
As a result, the overhead cost is more apparent for small-to-medium-sized transfers when the GET reply message is small enough, while the overhead is amortized in large-sized transfers.
Figure~\ref{fig:bw-results} also depicts the results of the previous works for comparison. Both TMD-MPI~\cite{tmd-mpi2} and THe GASNet~\cite{willenberg14-gasnet} achieved the peak bandwidth of 400 MB/s, while FSHMEM achieved 3813 MB/s, which is a 9.5$\times$ improvement. 
Compared to the one-sided MPI~\cite{ziavras-mpi}, FSHMEM achieves 26$\times$ bandwidth improvement.

\vspace{-0.05in}
\subsection{Communication Latency}

We measure FSHMEM's communication latency with the same experimental setting. From the time a command is given to the initiator FPGA, the PUT latency is measured until the message header is received in the remote FPGA, and the GET latency is measured until the reply message's header is returned to the initiator FPGA.
Table \ref{tab:lat} summarizes the latency measurement results of the various implementations for PUT and GET function. 
The FSHMEM's average latency for short messages (no payload) is measured at 0.21$\mu s$ and 0.45$\mu s$ for PUT and GET functions, while the average latency for long messages (payload size: 4 B to 2 MB) is measured at 0.35$\mu s$ and 0.59$\mu s$ for PUT and GET function, respectively.
The GET latency is by nature longer than that of PUT as the function requires two-way communication where a request is replied with the requested data.


\begin{table}[t]
    \centering
    \caption{Latency Comparison}
    \label{tab:lat}
    \def\arraystretch{1.3}%
    \begin{tabular}{ p{0.5\linewidth} c c }
        \hline
        Implementations & PUT ($\mu s$) & GET ($\mu s$) \\
        \hline
        TMD-MPI (inter-m2b) & \multicolumn{2}{c}{2} \\
        One-sided MPI & 0.36 & 0.62 \\
        THe GASNet (short message) & 0.17 & 0.35 \\
        THe GASNet (single word) & 0.29 & 0.47 \\
        FSHMEM (short message) & 0.21 & 0.45 \\
        FSHMEM (long message) & 0.35 & 0.59 \\
        \hline
    \end{tabular}
    \vspace{-0.2in}
\end{table}

Table~\ref{tab:lat} also shows a huge difference in latency between the TMD-MPI with two-sided communication and the other one-sided communication protocols including FSHMEM.
THe GASNet shows lower latency than FSHMEM through onboard wires. However, such channels are less scalable than FSHMEM's QSFP+ cables, commonly used in data centers.

\vspace{-0.05in}
\subsection{Comparison}

\begin{table}
    \centering
    \caption{Comparison with Prior Works}
    \vspace{-0.02in}
    \label{tab:comparison}
    \def\arraystretch{1.4}%
    \begin{tabular}{ p{0.14\linewidth} p{0.16\linewidth} p{0.14\linewidth} p{0.175\linewidth} p{0.145\linewidth} }
        \hline
        {} & TMD-MPI\cite{tmd-mpi2} & One-sided MPI\cite{ziavras-mpi} & THe GASNet\cite{willenberg14-gasnet} & This Work \\
        \hline
        FPGA & Xilinx XC5VLX110 & Xilinx XC2V6000 & Xilinx XC5VLX155T & Intel Stratix-10 \\
        Clock & 133.33 MHz & 50 MHz & 100 MHz & 250 MHz \\
        Data width & 32-bit & 32-bit & 32-bit & 128-bit \\
        Physical channel & Intel Front Side Bus & On-board wires & On-board wires & QSFP+ \\
        Max BW & 400 MB/s & 141 MB/s & 400 MB/s & 3813 MB/s \\
        Efficiency & 0.75 & 0.706 & 1.00 & 0.95 \\
        \hline
    \end{tabular}
    \vspace{-0.05in}
\end{table}

Table~\ref{tab:comparison} summarizes FSHMEM's implementation details compared to the previous implementations. 
FSHMEM achieves the highest communication bandwidth of 3813 MB/s with 95\% efficiency by utilizing a high-speed QSFP+ interface and lightweight GASNet core implementation.

\vspace{-0.05in}
\section{Case Study}
\vspace{-0.02in}

\begin{figure}[t]
 \centering
 \includegraphics[width=0.97\linewidth]{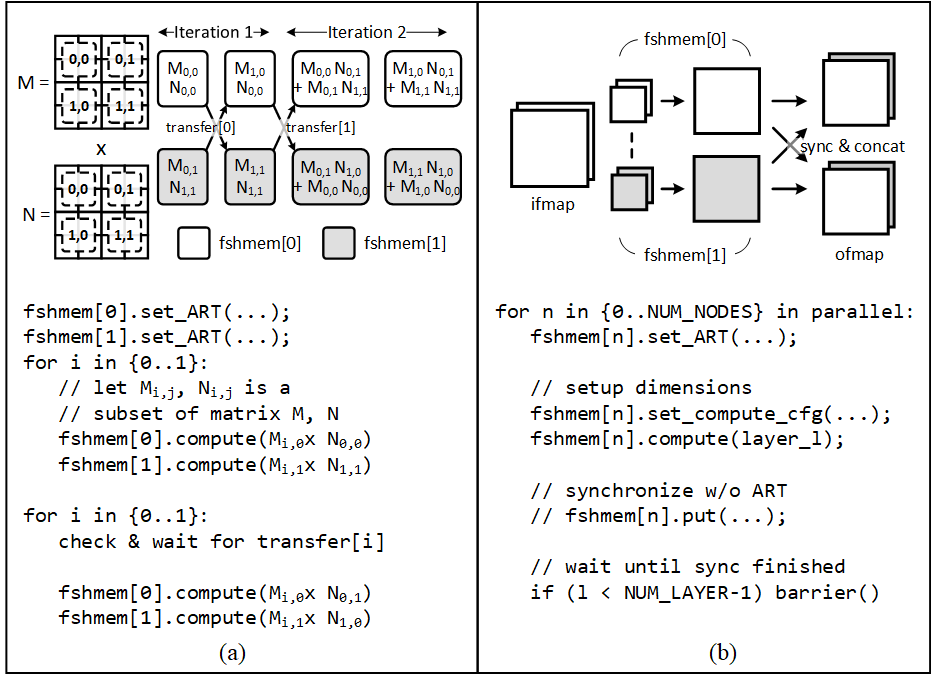}
 \caption{Parallel programs for (a) matrix multiplication and (b) convolution with their pseudo codes}
 \label{fig:app-pseudo-code}
 \vspace{-\baselineskip}
 \vspace{-0.02in}
\end{figure}

\begin{figure}[t]
 \vspace{-0.15in}
 \centering
 \includegraphics[width=0.97\linewidth]{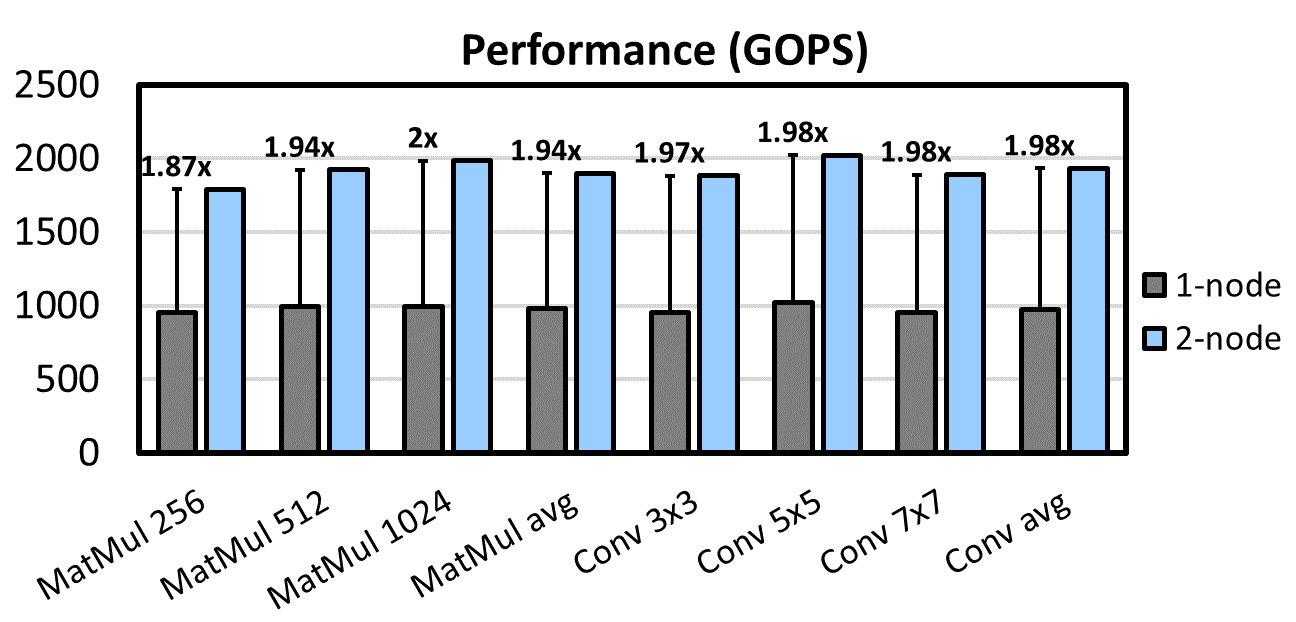}
 \caption{Performance results for matrix multiplication and convolution and the speedups achieved using 2 nodes}
 \label{fig:app-results}
 \vspace{-\baselineskip}
 \vspace{-0.05in}
\end{figure}

We conduct two case studies using FSHMEM to show its feasibility in parallel programming, especially for AI applications. We map a general matrix multiplication and convolution workload on the two FPGA nodes where each node integrates the Intel DLA with 16$\times$8 PEs.

Figure~\ref{fig:app-pseudo-code}(a) shows the implementation of parallel matrix multiplication ($M \times N$) using FSHMEM. Each of the two input matrices is partitioned into four sub-matrices, and the sub-matrices are split across the two FPGA nodes. 
The computation starts by iterating the row of matrix $M$ to multiply its sub-matrix with that of matrix $N$ (e.g., $N_{0,0}$, $N_{1,1}$), followed by exchanging the partial sum result to another node. After the first iteration, it checks if the first partial sum is transferred and does the same operation with the next set of matrix $N$ sub-matrices (e.g., $N_{0,1}$, $N_{1,0}$). Each node accumulates the partial sum to the previously transferred partial sum to get the final result. The result are stored across the two FPGAs like matrix $M$, where each FPGA holds sub-matrices of the same column.
Note that the command to transfer the partial sum is expressed by setting up the ART instead of explicitly using a PUT function in the pseudo code, allowing FSHMEM to send results simultaneously with computations to maximize the speedup.
We perform experiments on three different matrix sizes: 256$\times$256, 512$\times$512, and 1024$\times$1024.

For convolution, a widely used operation in convolutional neural networks (CNN), we split the weight kernels into two groups for parallel computation, as shown in Figure~\ref{fig:app-pseudo-code}(b).
After each convolution, both nodes must synchronize their results and concatenate them, producing a complete result in both nodes.
We use 64$\times$64 input feature maps for the experiments and vary the number and size of kernels: {256, 3$\times$3$\times$256}, {192, 5$\times$5$\times$192}, and {128, 7$\times$7$\times$128}.

Figure~\ref{fig:app-results} shows the experimental results on the two workloads comparing the single-node and two-node performance. For matrix multiplication, the single-node FPGA achieves an average of 979.4 GOPS, reaching 95.6\% of the theoretical maximum, while the two-node implementation achieves 1898.5 GOPS, which is a 1.94$\times$ performance gain.
We can see that the speedup increases as the matrix size increases because the longer accumulation in a larger matrix size gives more time to transfer the partial sum from one node to another.
The convolution operation's average performance gain is about 1.98$\times$ with 1931.3 GOPS. In general, convolution requires longer accumulation than matrix multiplication, resulting in a higher average speedup. However, none of the convolution results reach 2$\times$ speedup. One of the matrix multiplication results reaches 2$\times$ speedup as its algorithm hides the communication latency in-between the process, while the synchronization process in convolutions happens at the end of the process, causing an inevitable latency. Overall, all performances exceed the throughput of a single node by around 1.95$\times$ on average, suggesting a nearly linear speedup as the number of nodes increases.

\vspace{-0.05in}
\section{Conclusion and Future Work}
\vspace{-0.02in}
To conclude, we propose FSHMEM, a software/hardware framework for GASNet-enabled FPGA hardware acceleration infrastructure, by implementing the GASNet core and the supporting API in software.
We describe how this framework implements GASNet's AM functions in hardware and the software stack that enables high compatibility.
The benchmark results show that FSHMEM's bandwidth outperforms the prior works with competitive latency. Our case study on parallel matrix multiplication and convolution also shows great potential for scaling up.
For future work, we plan to build a scaled-up server that contains up to 8 FPGA acceleration cards and build an FSHMEM-based hardware infrastructure using it. 
We also plan to accelerate various machine learning models using the PGAS programming model for AI-enabled HPC.



\typeout{}
\bibliographystyle{IEEEtran}
\bibliography{ms}

\begin{thebibliography}{10}
\providecommand{\url}[1]{#1}
\csname url@samestyle\endcsname
\providecommand{\newblock}{\relax}
\providecommand{\bibinfo}[2]{#2}
\providecommand{\BIBentrySTDinterwordspacing}{\spaceskip=0pt\relax}
\providecommand{\BIBentryALTinterwordstretchfactor}{4}
\providecommand{\BIBentryALTinterwordspacing}{\spaceskip=\fontdimen2\font plus
\BIBentryALTinterwordstretchfactor\fontdimen3\font minus
  \fontdimen4\font\relax}
\providecommand{\BIBforeignlanguage}[2]{{%
\expandafter\ifx\csname l@#1\endcsname\relax
\typeout{** WARNING: IEEEtran.bst: No hyphenation pattern has been}%
\typeout{** loaded for the language `#1'. Using the pattern for}%
\typeout{** the default language instead.}%
\else
\language=\csname l@#1\endcsname
\fi
#2}}
\providecommand{\BIBdecl}{\relax}
\BIBdecl

\bibitem{shaw-molecular-hpc}
D.~E. Shaw \emph{et~al.}, ``Anton 2: Raising the bar for performance and
  programmability in a special-purpose molecular dynamics supercomputer,'' in
  \emph{SC '14: Proceedings of the International Conference for High
  Performance Computing, Networking, Storage and Analysis}, 2014, pp. 41--53.

\bibitem{jordan-weather-hpc}
\BIBentryALTinterwordspacing
J.~G. Powers \emph{et~al.}, ``The weather research and forecasting model:
  Overview, system efforts, and future directions,'' \emph{Bulletin of the
  American Meteorological Society}, vol.~98, no.~8, pp. 1717 -- 1737, 2017.
  [Online]. Available:
  \url{https://journals.ametsoc.org/view/journals/bams/98/8/bams-d-15-00308.1.xml}
\BIBentrySTDinterwordspacing

\bibitem{simpson-genomics-hpc}
\BIBentryALTinterwordspacing
J.~T. Simpson and R.~Durbin, ``{Efficient construction of an assembly string
  graph using the FM-index},'' \emph{Bioinformatics}, vol.~26, no.~12, pp.
  i367--i373, 06 2010. [Online]. Available:
  \url{https://doi.org/10.1093/bioinformatics/btq217}
\BIBentrySTDinterwordspacing

\bibitem{vinyals2016show}
O.~Vinyals, A.~Toshev, S.~Bengio, and D.~Erhan, ``Show and tell: Lessons
  learned from the 2015 mscoco image captioning challenge,'' \emph{IEEE
  Transactions on Pattern Analysis and Machine Intelligence}, vol.~39, no.~4,
  pp. 652--663, 2017.

\bibitem{cornia2020meshed}
M.~Cornia, M.~Stefanini, L.~Baraldi, and R.~Cucchiara, ``Meshed-memory
  transformer for image captioning,'' in \emph{Proceedings of the IEEE/CVF
  Conference on Computer Vision and Pattern Recognition}, 2020, pp.
  10\,578--10\,587.

\bibitem{Capes2017}
\BIBentryALTinterwordspacing
T.~Capes \emph{et~al.}, ``Siri on-device deep learning-guided unit selection
  text-to-speech system,'' in \emph{Proc. Interspeech 2017}, 2017, pp.
  4011--4015. [Online]. Available:
  \url{http://dx.doi.org/10.21437/Interspeech.2017-1798}
\BIBentrySTDinterwordspacing

\bibitem{Mehtab2021stock}
S.~Mehtab, J.~Sen, and A.~Dutta, \emph{Stock Price Prediction Using Machine
  Learning and LSTM-Based Deep Learning Models}, S.~M. Thampi, S.~Piramuthu,
  K.-C. Li, S.~Berretti, M.~Wozniak, and D.~Singh, Eds.\hskip 1em plus 0.5em
  minus 0.4em\relax Singapore: Springer Singapore, 2021.

\bibitem{tesla2020micro}
E.~Talpes \emph{et~al.}, ``Compute solution for tesla's full self-driving
  computer,'' \emph{IEEE Micro}, vol.~40, no.~2, pp. 25--35, 2020.

\bibitem{coates2013deep}
A.~Coates, B.~Huval, T.~Wang, D.~Wu, B.~Catanzaro, and N.~Andrew, ``Deep
  learning with cots hpc systems,'' in \emph{International conference on
  machine learning}.\hskip 1em plus 0.5em minus 0.4em\relax PMLR, 2013, pp.
  1337--1345.

\bibitem{kalamkar2020optimizing}
D.~Kalamkar, E.~Georganas, S.~Srinivasan, J.~Chen, M.~Shiryaev, and
  A.~Heinecke, ``Optimizing deep learning recommender systems training on cpu
  cluster architectures,'' in \emph{SC20: International Conference for High
  Performance Computing, Networking, Storage and Analysis}.\hskip 1em plus
  0.5em minus 0.4em\relax IEEE, 2020, pp. 1--15.

\bibitem{monroe2020fugaku}
\BIBentryALTinterwordspacing
D.~Monroe, ``Fugaku takes the lead,'' \emph{Commun. ACM}, vol.~64, no.~1, p.
  16–18, dec 2020. [Online]. Available: \url{https://doi.org/10.1145/3433954}
\BIBentrySTDinterwordspacing

\bibitem{hines2018stepping}
J.~Hines, ``Stepping up to summit,'' \emph{Computing in science \&
  engineering}, vol.~20, no.~2, pp. 78--82, 2018.

\bibitem{mpi40}
\BIBentryALTinterwordspacing
{Message Passing Interface Forum}, \emph{{MPI}: A Message-Passing Interface
  Standard Version 4.0}, Jun. 2021. [Online]. Available:
  \url{https://www.mpi-forum.org/docs/mpi-4.0/mpi40-report.pdf}
\BIBentrySTDinterwordspacing

\bibitem{yelick2007prod}
\BIBentryALTinterwordspacing
K.~Yelic \emph{et~al.}, ``Productivity and performance using partitioned global
  address space languages,'' in \emph{Proceedings of the 2007 International
  Workshop on Parallel Symbolic Computation}, ser. PASCO '07.\hskip 1em plus
  0.5em minus 0.4em\relax New York, NY, USA: Association for Computing
  Machinery, 2007, p. 24–32. [Online]. Available:
  \url{https://doi.org/10.1145/1278177.1278183}
\BIBentrySTDinterwordspacing

\bibitem{brainwave2018}
\BIBentryALTinterwordspacing
E.~Chung \emph{et~al.}, ``Serving dnns in real time at datacenter scale with
  project brainwave,'' \emph{IEEE Micro}, vol.~38, pp. 8--20, March 2018.
  [Online]. Available:
  \url{https://www.microsoft.com/en-us/research/publication/serving-dnns-real-time-datacenter-scale-project-brainwave/}
\BIBentrySTDinterwordspacing

\bibitem{mango}
R.~Tornero-Gavil{\'a}, J.~Flich~Cardo, J.~M. Mart{\'\i}nez~Mart{\'\i}nez,
  T.~Picornell-Sanjuan, and V.~Scotti, ``The mango process for designing and
  programming multi-accelerator multi-fpga systems,'' in \emph{Fourth
  International Workshop on Heterogeneous High-Performance Reconfigurable
  Computing (H2RC'18)}.\hskip 1em plus 0.5em minus 0.4em\relax ACM, 2018.

\bibitem{cygnus}
B.~Taisuke \emph{et~al.}, ``Cygnus: A multi-hybrid supercomputing platform with
  gpus and fpgas,''
  \textsc{url:}~\url{https://www.isc-hpc.com/agenda2019/conferences/isc_hpc/assets/2019/posters/proj138.pdf},
  June 2019, poster on ISC2019.

\bibitem{xiong20-collectives}
Q.~Xiong, C.~Yang, P.~Haghi, A.~Skjellum, and M.~Herbordt, ``Accelerating mpi
  collectives with fpgas in the network and novel communicator support,'' in
  \emph{2020 IEEE 28th Annual International Symposium on Field-Programmable
  Custom Computing Machines (FCCM)}, 2020, pp. 215--215.

\bibitem{tmd-mpi}
M.~Saldana and P.~Chow, ``Tmd-mpi: An mpi implementation for multiple
  processors across multiple fpgas,'' in \emph{2006 International Conference on
  Field Programmable Logic and Applications}, 2006, pp. 1--6.

\bibitem{patel06-scalable}
A.~Patel \emph{et~al.}, ``A scalable fpga-based multiprocessor,'' in \emph{2006
  14th Annual IEEE Symposium on Field-Programmable Custom Computing Machines},
  2006, pp. 111--120.

\bibitem{ly09-challenge}
D.~L. Ly, M.~Saldaña, and P.~Chow, ``The challenges of using an embedded mpi
  for hardware-based processing nodes,'' in \emph{2009 International Conference
  on Field-Programmable Technology}, 2009, pp. 120--127.

\bibitem{axel}
\BIBentryALTinterwordspacing
K.~H. Tsoi and W.~Luk, ``Axel: A heterogeneous cluster with fpgas and gpus,''
  in \emph{Proceedings of the 18th Annual ACM/SIGDA International Symposium on
  Field Programmable Gate Arrays}, ser. FPGA '10.\hskip 1em plus 0.5em minus
  0.4em\relax New York, NY, USA: Association for Computing Machinery, 2010, p.
  115–124. [Online]. Available: \url{https://doi.org/10.1145/1723112.1723134}
\BIBentrySTDinterwordspacing

\bibitem{willenberg14-gasnet}
\BIBentryALTinterwordspacing
R.~Willenberg and P.~Chow, ``A heterogeneous gasnet implementation for
  fpga-accelerated computing,'' in \emph{Proceedings of the 8th International
  Conference on Partitioned Global Address Space Programming Models}, ser. PGAS
  '14.\hskip 1em plus 0.5em minus 0.4em\relax New York, NY, USA: Association
  for Computing Machinery, 2014. [Online]. Available:
  \url{https://doi.org/10.1145/2676870.2676885}
\BIBentrySTDinterwordspacing

\bibitem{gasnet_1.8.1}
D.~Bonachea and P.~Hargrove, ``Gasnet specification, v1.8.1,'' Lawrence
  Berkeley National Laboratory, Tech. Rep., 8 2017.

\bibitem{am}
T.~von Eicken, D.~E. Culler, S.~C. Goldstein, and K.~E. Schauser, ``Active
  messages,'' \emph{ACM SIGARCH Computer Architecture News}, vol.~20, pp.
  256--266, 5 1992.

\bibitem{bonachea2018gasnet}
D.~Bonachea and P.~H. Hargrove, ``Gasnet-ex: A high-performance, portable
  communication library for exascale,'' in \emph{International Workshop on
  Languages and Compilers for Parallel Computing}.\hskip 1em plus 0.5em minus
  0.4em\relax Springer, 2018, pp. 138--158.

\bibitem{tmd-mpi2}
\BIBentryALTinterwordspacing
M.~Salda\~{n}a \emph{et~al.}, ``Mpi as a programming model for high-performance
  reconfigurable computers,'' \emph{ACM Trans. Reconfigurable Technol. Syst.},
  vol.~3, no.~4, nov 2010. [Online]. Available:
  \url{https://doi.org/10.1145/1862648.1862652}
\BIBentrySTDinterwordspacing

\bibitem{ziavras-mpi}
\BIBentryALTinterwordspacing
S.~G. Ziavras, A.~V. Gerbessiotis, and R.~Bafna, ``Coprocessor design to
  support mpi primitives in configurable multiprocessors,'' \emph{Integration},
  vol.~40, no.~3, pp. 235--252, 2007. [Online]. Available:
  \url{https://www.sciencedirect.com/science/article/pii/S0167926005000519}
\BIBentrySTDinterwordspacing

\bibitem{haghi20fpga}
P.~Haghi \emph{et~al.}, ``Fpgas in the network and novel communicator support
  accelerate mpi collectives,'' in \emph{2020 IEEE High Performance Extreme
  Computing Conference (HPEC)}, 2020, pp. 1--10.

\bibitem{intel-dla}
M.~S. {Abdelfattah} \emph{et~al.}, ``Dla: Compiler and fpga overlay for neural
  network inference acceleration,'' in \emph{2018 28th International Conference
  on Field Programmable Logic and Applications (FPL)}, 2018, pp. 411--418.

\bibitem{intel-opae}
\BIBentryALTinterwordspacing
{Intel Corporation}, ``Open programmable acceleration engine.'' [Online].
  Available: \url{https://01.org/opae}
\BIBentrySTDinterwordspacing

\bibitem{xilinx-xrt}
\BIBentryALTinterwordspacing
{Xilinx, Inc.}, ``Xilinx runtime.'' [Online]. Available:
  \url{https://github.com/Xilinx/XRT}
\BIBentrySTDinterwordspacing

\end{thebibliography}

\end{document}